\documentclass[12pt,a4paper]{article}
\usepackage{epsf}
\usepackage{epsfig}
\usepackage{feynmp}
\usepackage{rotating}

\def\lt{$<$}

\def\be{\begin{equation}}
\def\ee{\end{equation}}
\def\bea{\begin{eqnarray}}
\def\eea{\end{eqnarray}}
\def\beas{\begin{eqnarray*}}
\def\eeas{\end{eqnarray*}}

\def\a_la_ligne{}

\def\tanb{\mbox{$\tan\beta$}}

\def\infb{\mbox{$\hbox{fb}^{-1}$}}
\def\invfb{\mbox{$\hbox{fb}^{-1}$}}

\def\Gcs{\hbox{GeV}/\mbox{$\mathrm{c^2}$}}
\def\Gc{\hbox{GeV}/\mbox{$\mathrm{c}$}}
\def\Tcs{\hbox{TeV}/\mbox{$\mathrm{c^2}$}}

\def\Z{\mbox{$\hbox{\rm Z}$}}

\def\W{\mbox{$\hbox{\rm W}$}}

\def\A{\mbox{$\hbox{\rm A}$}}
\def\H{\mbox{$\hbox{\rm H}$}}

\def\WpWm{\mbox{$\rm \W^+\W^-$}}

\def\h{\mbox{$ \hbox{\rm h}$}}

\def\goes{\mbox{$\rightarrow$}}

\def\epem{\mbox{$\mathrm{e^+ e^-}$}}

\def\lplm{\mbox{$\ell^+\ell^-$}}
\def\nnbar{\mbox{$\nu\bar\nu$}}

\def\qqbar{\mbox{${\mathrm {q\bar{q}}}$}}

\def\ttbar{\mbox{${\mathrm {t\bar{t}}}$}}

\def\bbbar{\mbox{${\mathrm {b\bar{b}}}$}}

\def\sqrtofs{\mbox{$\rm \sqrt s $}}

\def\mh{\mbox{$m_{\mathrm h}$}}

\def\mhsq{\mbox{$m^2_{\mathrm h}$}}

\def\mA{\mbox{$m_{\mathrm A}$}}

\def\mZ{\mbox{$m_{\mathrm Z}$}}

\def\mstop{\mbox{$m_{\tilde{\mathrm{t}}}$}}

\def\ie{{\it{i.e.}}}

\newcommand{\capstyl}{\small \sf }

\newcommand{\source}[1]{} 

\def\fom{$\rm s/\sqrt{b}$ }

\newcommand{\tento}[1]{$\mathrm{10^{#1}}$ }

\def\SM{standard model}

\def\lumi{${\cal L}$} 
\def\lumisim{${\cal L}_{\rm sim} $} 
\def\Mhemi{$\rm M_{hemi}$} 
\def\pztot{$\rm P_z^{tot}$} 
 
\def\Zgamma{\Z($\gamma$)} 
\def\ycut{$\rm y_{cut}$}

\def\trilin{$\lambda_{\h\h\h}$}

\def\LPC{$^{(a)}$}
\def\DAPNIA{$^{(b)}$}
\def\fb{fb}
\def\brecoil{$\rm {\cal{B}}^{recoil}$}
\def\capstyl{\small \sl } 
\def\invab{$\rm ab^{-1}$} 

\begin{document}
%
%
%
\begin{titlepage}
\setlength{\topmargin}{0.5cm}
\setlength{\oddsidemargin}{-0.2cm}

\title{Higgs self coupling measurement in e$^+$e$^-$ collisions at center-of-mass energy of 500~GeV}

\author{
C.~Castanier\LPC, P.~Gay\LPC, P.~Lutz\DAPNIA, J.~Orloff\LPC \\ 
\begin{tabular}{c} 
 \cr \cr  
\LPC\ {\it \footnotesize Laboratoire de Physique Corpusculaire, Univ. B. Pascal/$ IN^2P^3$-CNRS,} \cr  
{\it  \footnotesize 24 Av. des Landais, F-63177 Aubi\`ere Cedex, France} \cr  
\cr 
\DAPNIA\ {\it  \footnotesize CEA/DAPNIA, Service de Physique de particules,} \cr 
{\it  \footnotesize CE-Saclay, F-91191 Gif-sur-Yvette Cedex, France} \cr \end{tabular} 
}
\date{\mbox{ }}
\maketitle
\thispagestyle{empty}

\begin{picture}(160,1)
\put(117, 134){\parbox[t]{45mm}{{\tt LC-PHSM-2000-061 \\ hep-ex/0101028}}}
\end{picture}

\begin{abstract}
\vspace{.5cm}
Feasibility of the measurement of the trilinear self-couplings of 
the Higgs boson is studied. Such a measurement would experimentally 
determine the structure of the Higgs potential. Full hadronic and 
semi-leptonic final states of the double-Higgs strahlung have been 
investigated.
\end{abstract}
\vfill



\end{titlepage}

%
\cleardoublepage 
\newpage 
\setlength{\topmargin}{-1cm}
\setlength{\oddsidemargin}{-0.5cm}

\newpage
\pagestyle{plain}
\setcounter{page}{1}
%
%
\pagenumbering{arabic}
\normalsize

\setlength{\textheight}{23cm}
\setlength{\textwidth}{17cm}
\unitlength 1mm
\setlength{\topmargin}{-1cm}
\setlength{\oddsidemargin}{-0.5cm}

\section{Introduction}
\label{intro} 

In the framework of the standard model, the generation of mass occurs
through the Higgs mechanism. This mechanism assumes a Higgs potential,
V($\Phi$), which behaves as \mbox{V($\Phi$)={$\boldmath{\lambda}$}($\Phi^2$-$\frac{1}{2}${\it v}$^2$)$^2$}, where
$\phi$ is an isodoublet scalar field, and {\it v}$\sim$246 GeV is the
vacuum expectation value of its neutral component. A determination of
the Higgs boson mass, which satisfies \mhsq=4$\lambda${\it v}$^2$ at
tree level in the \SM{}, will provide an indirect information about
the Higgs potential and its self-coupling $\lambda$. The measurement
of the trilinear self-coupling $\lambda_{\h\h\h}$=$\frac{6}{\sqrt{2}}$
$\lambda${\it v} offers an independent determination of the Higgs
potential shape.  In case of disagreement, the comparison between
these two measurements would give an interesting clue about new
physics, like super-symmetry for instance.
Anyway, the reconstruction of the Higgs potential is an
essential step in the experimental validation of the Higgs mechanism
that lies at the core of the \SM{}.

The aim of our study is to demonstrate how to measure
the trilinear self-coupling $\lambda_{\h\h\h}$ in \epem{} collisions
at center-of-mass energies delivered by the future Linear Collider.
The note is organized as follows~: the framework of the study is
defined in Section~\ref{signal}. The Monte Carlo simulation is
reported in Section~\ref{mc}. Analyses and results are described in
Section~\ref{analysis} and~\ref{results}. 

\section{Signal} 
\label{signal} 

\begin{figure}[htb]
\begin{center} 
\vspace{1.cm}
\vspace{.7cm}
{ \footnotesize
\unitlength1mm
\hspace{10mm}
\begin{fmfshrink}{0.7}
\begin{fmfgraph*}(24,12)
  \fmfstraight
  \fmfleftn{i}{3} \fmfrightn{o}{3}
  \fmf{fermion}{i1,v1,i3}
  \fmflabel{$e^-$}{i1} \fmflabel{$e^+$}{i3}
  \fmf{boson,lab=$Z$,lab.s=left,tens=3/2}{v1,v2}
  \fmf{boson}{v2,o3} \fmflabel{$Z$}{o3}
  \fmf{phantom}{v2,o1}
  \fmffreeze
  \fmf{dashes,lab=$H$,lab.s=right}{v2,v3} \fmf{dashes}{v3,o1}
  \fmffreeze
  \fmf{dashes}{v3,o2} 
  \fmflabel{$H$}{o2} \fmflabel{$H$}{o1}
  \fmfdot{v3}
\end{fmfgraph*}
\hspace{15mm}
\begin{fmfgraph*}(24,12)
  \fmfstraight
  \fmfleftn{i}{3} \fmfrightn{o}{3}
  \fmf{fermion}{i1,v1,i3}
  \fmf{boson,lab=$Z$,lab.s=left,tens=3/2}{v1,v2}
  \fmf{dashes}{v2,o1} \fmflabel{$H$}{o1}
  \fmf{phantom}{v2,o3}
  \fmffreeze
  \fmf{boson}{v2,v3,o3} \fmflabel{$Z$}{o3}
  \fmffreeze
  \fmf{dashes}{v3,o2} 
  \fmflabel{$H$}{o2} \fmflabel{$H$}{o1}
\end{fmfgraph*}
\hspace{15mm}
\begin{fmfgraph*}(24,12)
  \fmfstraight
  \fmfleftn{i}{3} \fmfrightn{o}{3}
  \fmf{fermion}{i1,v1,i3}
  \fmf{boson,lab=$Z$,lab.s=left,tens=3/2}{v1,v2}
  \fmf{dashes}{v2,o1} \fmflabel{$H$}{o1}
  \fmf{dashes}{v2,o2} \fmflabel{$H$}{o2}
  \fmf{boson}{v2,o3} \fmflabel{$Z$}{o3}
\end{fmfgraph*}
\end{fmfshrink}
}
\end{center} 
\caption{\capstyl Feynman diagrams involved in the \epem\goes\h\h\Z\ cross-section via the double Higgs-strahlung.\label{fda}}
\end{figure}
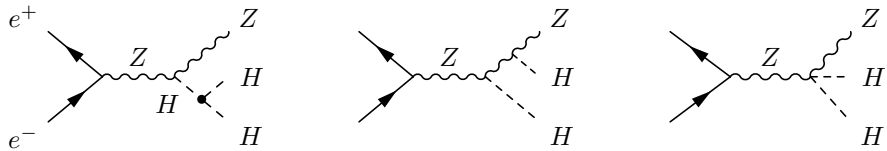

The trilinear Higgs self-coupling, $\lambda_{\h\h\h}$, could be
extracted from the measurement of the cross-section of each of the
following processes~: double Higgs-strahlung ($e^+e^-\to Z\h\h$) or
$WW$ double-Higgs fusion ($e^+e^-\to \bar\nu_e\nu_e \h\h$)~\cite{djouadi}.
Figure~\ref{fda} indicates the major Feynman diagrams
involved in the first process.

\begin{fmffile}{fd}
\end{fmffile}

For light Higgs boson masses, the Higgs boson decays predominantly in
a \bbbar{} pair. When the \Z{} decays in a lepton pair, the final
state is \h\h\Z\goes\bbbar\bbbar\lplm. With a high b
content and two leptons, this topology produces an easy signature but
represents only $\sim$ 8\%{} of the total final state. By contrast,
the \h\h\Z\goes\bbbar\bbbar\qqbar{} final state benefits from a high
statistics with $\sim$ 60\%{} of the final states but requires a more
complicated analysis.

\begin{figure}[htbp]
\begin{center}
\begin{picture}(140,100)
\put(30,0){\epsfxsize110mm\epsfbox{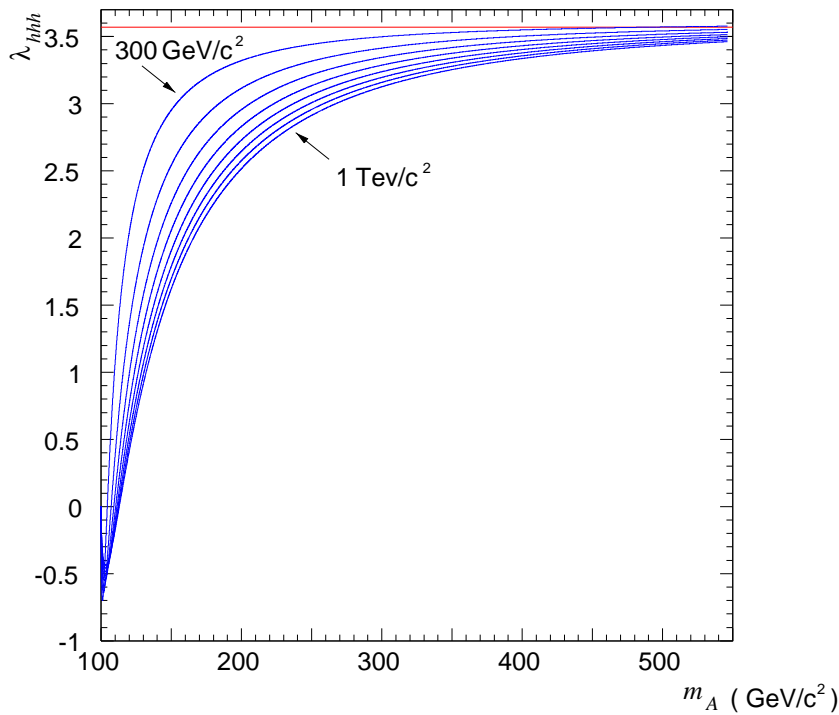}}
\end{picture}
\caption{\capstyl MSSM trilinear Higgs self-coupling (\trilin) as a function 
  of the mass of the pseudo-scalar Higgs boson mass (\mA{}) and different assumptions of \mstop\ from 300~\Gcs\ to 1~\Tcs\ with a 100~\Gcs\ step size; the horizontal line corresponds to the \SM\ value of \trilin. A 100~\Gcs\ Higgs boson mass has been assumed. 
\label{mssm}}
\end{center}
\end{figure}

In the minimal super-symmetric extension (MSSM) of the \SM, the Higgs
sector is richer with three neutral Higgs bosons (\h, \H{} and \A).
The Higgs potential is therefore more intricate and further trilinear
self couplings are possible beyond $\lambda_{\h\h\h}$~:
$\lambda_{\h\h\H}$, $\lambda_{\H\H\H}$, $\lambda_{\h\A\A}$, and
$\lambda_{\H\A\A}$. At the tree level, these depend on two parameters
(\tanb{} and \mA{} for instance). The radiative corrections are driven
mainly by the top quark, so that the stop mass effectively constitutes
a third one.
Studies performed in the parameter space of the MSSM are described in
Ref.~\cite{djouadi,osland}. Here, let us simply recall on
Fig.~\ref{mssm} the sensitivity of \trilin{} as a function of \mA{}
with different \mstop{} assumptions. For fixed \mh{}, the
value of $\lambda_{\h\h\h}$ reaches its \SM{} value for large \mA{}
(decoupling limit), and the dependence on \mstop{} only shows up for
smaller values of \mA{}. However, the topologies obtained will be
similar to the \SM{} ones, except for particular parameter choices
where the Higgs boson decay into a \bbbar{} pair is suppressed.  Thus,
the experimental study we develop in this note, though a priori
restricted to the \SM{} framework, is also relevant for the MSSM.

If the mass of the Higgs boson is assumed to be light ($\sim$ 120~\Gcs)
then the best center-of-mass energy is around 500~GeV~\cite{djouadi}
and the double Higgs-strahlung process cross-section is one order of
magnitude greater than the \W\W\ fusion one. The study presented here
will be performed for \mh=120~\Gcs{} and \sqrtofs=500~GeV and it is
restricted to the \SM\ framework.

Nevertheless, with \mh=120~\Gcs\ and \sqrtofs=500~GeV the
\epem\goes\h\h\Z\ cross-section is still very tiny with a 0.18~\fb\ 
value. Namely, for an expected integrated luminosity of 500~\invfb,
only 93 signal events are produced. The difficulty is increased by the
the huge background sources, as detailed in next Section.

\section{Simulation}
\label{mc} 
The signal event samples have been simulated with the {\tt GRACE}
generator~\cite{grace}. The background sources are either two fermions
final state processes as \epem\goes\Z$\gamma$, or four fermions
processes like \epem\goes\WpWm, \Z\Z, \W e$\nu$, \Z ee, \h\Z.  The
latter has been simulated with {\tt HZHA} generator~\cite{hzha} while
{\tt PYTHIA}~\cite{pythia} was used for the others processes.  Six
fermions final state processes like \epem\goes\W\W\Z{} or \Z\Z\Z{} are
also background sources and {\tt GRACE} was used to simulate them.
Table~\ref{mctab} summaries the size of the samples used and the
cross-section of each process.  Since \epem\goes\ttbar{} and
\epem\goes\W tb will yield major contributions to the
background, an emphasis has been performed and a simulated luminosity
greater than 2000\infb{} was generated for such processes as well as
for six fermions processes while for the others a simulated luminosity
greater than 500\infb{} was used as reported in Table~\ref{mctab}. Even 
if the cross-section is small, \epem\goes\ttbar\h\ process is also 
a potential source of background and this process was considered.  \\ 

\begin{table} 
\begin{center} 
\begin{tabular}{|ll|c|c|c|c|}
\hline
& { process} &  { $N_{gen}$ }& {$ \sigma$ (fb) }& { generator }& { \lumisim (\infb)} \cr
\hline
\multicolumn{6}{l}{  Signal}  \cr 
\hline
& \h\h\Z\ (Z\goes\qqbar)        & 11k& 0.13 &   {\tt GRACE} & 84.10$^3$\cr 
& \h\h\Z\ (Z\goes\lplm)         & 5k& 0.02 &   {\tt GRACE} & 270.10$^3$\cr 
\hline 
\multicolumn{6}{l}{  2 fermions} \cr 
\hline
&\Z$\gamma$                    & 4.2M&8200.&   {\tt PYTHIA}  & 514.\cr 
&\Z$\gamma$\goes\ttbar$\gamma$ & 150k&550.&   {\tt PYTHIA}  & 2145.\cr 
\hline 
\multicolumn{6}{l}{  4 fermions} \cr 
\hline
&\W\W                          & 3.9M& 7700.&    {\tt PYTHIA}  & 509.\cr 
&\W\W\goes\W tb                & 17k& 16.8&    {\tt PYTHIA}  & 12.10$^3$\cr 
&\Z\Z                          & 300k & 550.&   {\tt PYTHIA}  & 545.\cr
&\W e$\nu$                     & 2.6M&5300.&    {\tt PYTHIA}  & 502.\cr 
&\Z ee                         & 3.7M&7400.&   {\tt PYTHIA}  & 504.\cr 
&\h\Z                          & 35k&70.5&   {\tt HZHA} & 1631.\cr 
&\ttbar\h, \h\Z\goes\ttbar\h   & 3k &0.4 &   {\tt GRACE}& 7500. \cr 
\hline 
\multicolumn{6}{l}{  6 fermions}  \cr 
\hline
&\W\W\Z\ (Z\goes\qqbar)        & 21k&19.8&   {\tt GRACE} & 3383.\cr 
&\W\W\Z\ (Z\goes\lplm)         & 8.6k&2.8&   {\tt GRACE} & 10.10$^3$.\cr
&\Z\Z\Z\ (Z\goes\qqbar)        & 6k &0.53&   {\tt GRACE} & 30.10$^3$.\cr
&\Z\Z\Z\ (Z\goes\lplm\ \nnbar) & 9.5k &1.01& {\tt GRACE} & 28.10$^3$.\cr
\hline
\end{tabular}
\caption[]{\capstyl Cross-sections for signal and background processes, Monte Carlo statistics 
and simulated luminosity (\lumisim).} \label{mctab}
\end{center}
\end{table}

The detector simulation was performed with a Parametric Monte Carlo~\cite{simdet}. The tracking system, immersed in a 3 tesla magnetic field, consists on a vertex detector (VDET) surrounding the beam tube at $\sim$ one centimeter radius followed by a time projection chamber (TPC). The VDET impact parameter resolution is assumed to be $\sim$5$\rm \mu m$ in both rz and $\rm r\phi$ views. Only charged particles with a transverse momentum greater than 0.2~\Gc{} are reconstructed. The tracking efficiency is assumed to be 99.7\%. The tracking system is complemented with a forward tracker and muon chambers. The b-tagging is performed through a parametrisation derived from full reconstruction~\cite{rh}.

At low angle, a luminometer (LCAL) enforces the hermeticity of the detector. 
Energy and direction of the photons and neutral hadrons 
are measured thanks to the electromagnetic (ECAL) and hadronic (HCAL) calorimeters. The intrinsic energy resolutions taken into account are 
 $\Delta E/E$=10.2\%/$\sqrt{E(GeV)}$ and  $\Delta E/E$=40.5\%/$\sqrt{E(GeV)}$  respectively for ECAL/LCAL and HCAL. The energy threshold below which no neutral particles is reconstructed in calorimeters are 200~MeV and 500~MeV for ECAL and HCAL respectively. Following studies performed with full simulation of high granularity calorimeters~\cite{eflow}, a jet energy resolution of 40\%/$\sqrt{E_{jet}}$ could be achieved. Such a resolution has been assumed in the analysis described hereafter.

\section{Analyses} 
\label{analysis} 
If an isolated track is allowed to form a jet, both final state, \h\h\qqbar{} and \h\h\lplm{} are characterized by six jets, and the hadronic system is characterized by a high b content. 

The analysis works in three steps. The preselection mainly based on event
shape variables consists in removing most of the two- and
four-fermions final-state contributions of the background. In a second
step the b-content as well as the di-jets reconstructed masses are
used to reject the three-bosons final-state contribution. Finally, all
the relevant informations are combined with a multivariable method.

\begin{figure}[htbp]
\begin{center}
\begin{picture}(160,70)
\put(0,0){\epsfxsize150mm\epsfbox{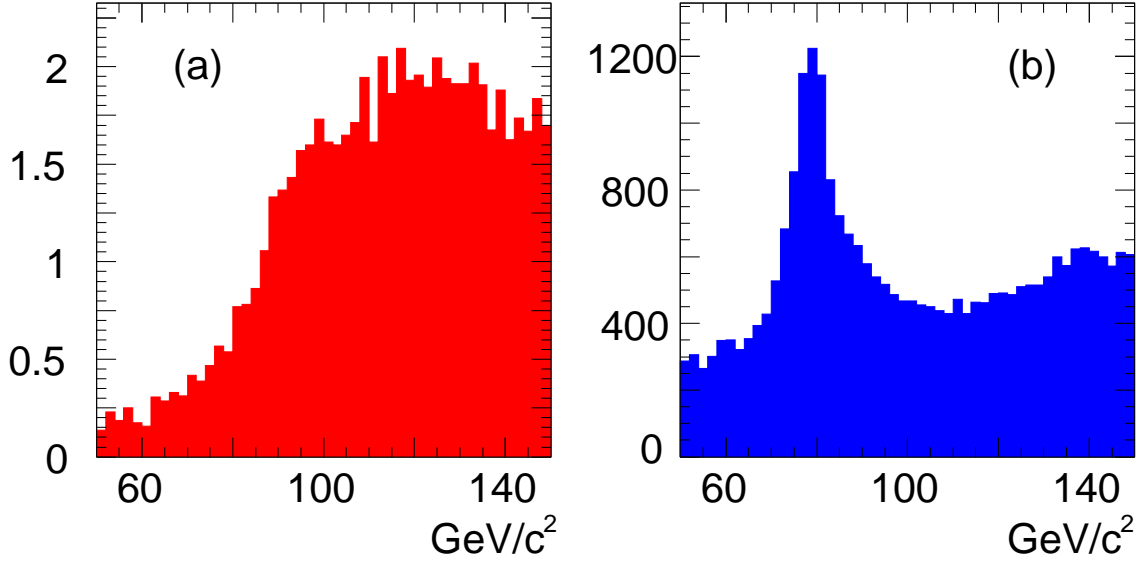}}
\end{picture}
\caption{\capstyl Distribution of the invariant mass of each hemisphere (\Mhemi) for (a) signal and (b) background processes. 
\label{hemi}}
\end{center}
\end{figure}

The preselection consists in rejecting events with a charged
multiplicity less than 30.  The thrust value has to be less than 0.9
and the direction of the thrust with respect to beam axis should
verify $\rm cos(\theta_{thrust}) < $ 0.9 .  In order to remove
\epem\goes\Zgamma, the longitudinal component of the total momentum of
the event (\pztot) have to satisfy \pztot\lt{} 50~\Gc.  The event is
divided in two hemispheres with respect to the thrust axis and the
invariant mass of each hemisphere (\Mhemi) must exceed 90~\Gcs{}.
Figure~\ref{hemi} illustrates the discrimination between the signal
and the background processes achieved by this variable.

The event is clustered with the {\tt DURHAM} scheme.  The variation of
the \ycut{} is stopped when the transition between six and seven jets
occurs and the event is thus forced into a six jets topology.  To
eliminate \epem\goes\Zgamma{} with a photon emitted in the sensitive
region of the ECAL, a fraction of electromagnetic energy with respect
to the jet energy less than 80\% is requested for all the jets.

The numbers of events expected from background processes at the preselection 
level are reported in Table~\ref{res}. The  contribution from \epem\goes\Z$ee$ or \W$\rm e \nu$ processes is canceled. Around 48.\tento{3} background events are expected with a luminosity of 500~\invfb{} while $\sim$42 and  $\sim$7 signal events are expected for \h\h\qqbar{} and \h\h\lplm{} final state respectively when \mh=120\Gcs{} is assumed. Major sources of background are processes with at least one top quark in the final state. In fact, 99\% of the contribution from the 
\epem\goes\Zgamma{} process is coming from \Z\goes\ttbar\ and 80\% of the contribution from \epem\goes\WpWm{} process is \epem\goes\W tb. The remaining part of those background processes mainly consists in six fermions final state with two or one b quarks. However, even if the background contribution is largely reduced, the remaining part is three order of magnitude greater than the signal.

\begin{figure}[htbp]
\begin{center}
\begin{picture}(140,75)
\put(-10,0){\epsfxsize170mm\epsfbox{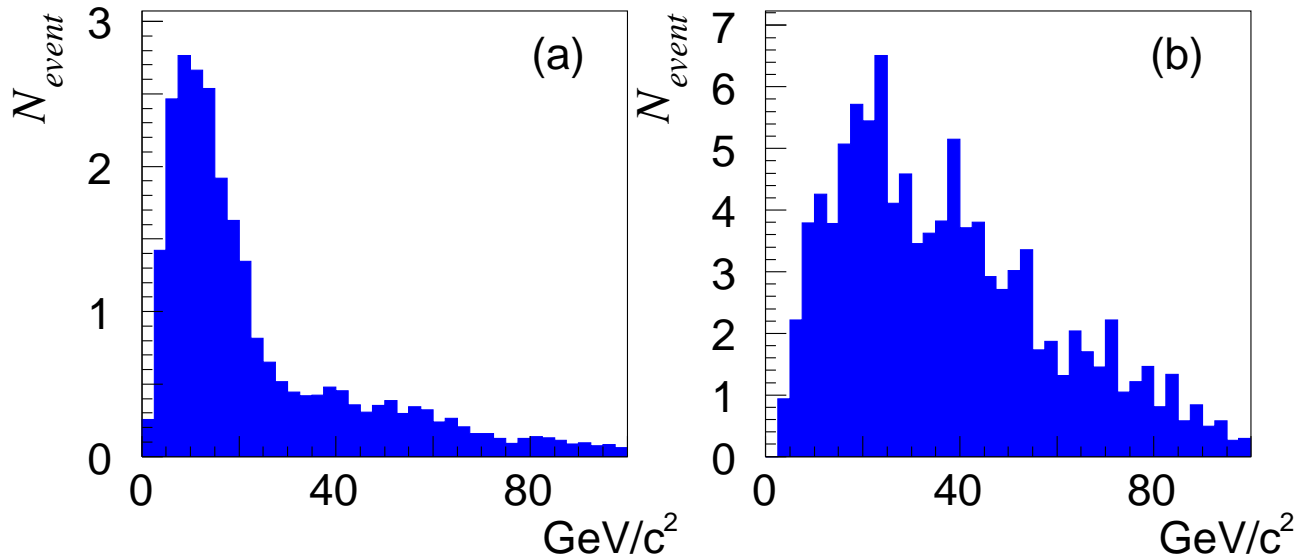}}
\end{picture}
\caption{\capstyl Distribution of the variable DIST (defined in text) for (a) signal and (b) background processes. 
\label{dist}}
\end{center}
\end{figure}

Thanks to the jet energy flow resolution, a direct use of the
reconstructed di-jets masses is applied. Among all the di-jets masses
in the event, the so-called $m_{56}$ mass minimizes the difference
with a \mZ\ hypothesis.  The b content of the system recoiling to the
\Z\ (\brecoil) is an excellent variable to reject contribution from
background processes.  In a first stage, a loose cut is applied~: at
least one b jet among the recoiling jets system (\brecoil $>$1) is
requested. Among the four jets recoiling to the Z (\ie\ $m_{56}$), the
di-jets masses $m_{12}$ and $m_{34}$ are defined by minimizing
$\|m_{12}$-$m_{34}\|$ while having both $m_{12}$ and $m_{34}$ larger
than 100~\Gcs. The three di-jets masses are combined in a simple way
in order to form the distance
DIST=$\sqrt((m_{12}-\mh)^2+(m_{34}-\mh)^2+(m_{56} -\mZ)^2)$, where
\mh\ and \mZ\ are the Higgs boson and Z masses.  The distribution of
this variable DIST is displayed on Fig.~\ref{dist} when a tightened
cut on \brecoil\ is applied (\brecoil$>$2).

The contribution from the various background sources are reported in Table~\ref{res} for the loose (\brecoil$>$1) and tightened (\brecoil$>$2) cuts. The signal efficiencies are  43\% and 35\% respectively. Most of the background is coming from \epem\goes\W tb and \epem\goes\ttbar\ processes while the six fermions final state contribution is due to \epem\goes\Z\Z\Z\ with relatively small level with respect to the previous contributions. The figure of merit defined as s/$\rm \sqrt{b}$ increases from 0.22 at the preselection level up to 3, with an integrated luminosity of 500~\invfb.

\begin{table}[hbt]
\begin{center} 
\begin{tabular}{|l|c|cc|c|}
\hline 
process & preselection             & b-content & b-content  &  NNet \cr 
        &                          & \brecoil$>$1      & \brecoil$>$2       &   $>$0    \cr
\hline
\h\h\qqbar        & 41.4 & 34.   &  27.1 &   27.5    \cr 
\h\h\lplm         & 6.7  & 6.2   &  5.1   &   6.4  \cr 
 total \h\h\Z & 49.1 & 40.2   &  32.2  &  33.9   \cr 
\hline
 \W\W &2114.           &233.       & 74.3 &32.            \cr  
\Z$\gamma$  &44938.          &116.       & 34. &24.          \cr 
\Z\Z        &484.            &7.4        & 0. &0.          \cr 
\W\W\Z\     &331.            &0.6        & 0.  &0.14           \cr
\Z\Z\Z      &56.6            &19.         & 9.  &8.4          \cr 
\h\Z        &174.            &0.         & 0.  &0.         \cr
\ttbar\h        &3.           &0.         & 0.  &0.         \cr
{ total bkg.}& { 48089.} &{ 376.}  & { 117.4} & { 64.3}\cr 
\hline
 s/b  & 0.1\%&  11\% &  27\%  & 53\% \cr 
 \fom & 0.22   & 2.  &  3. & 4.2 \cr \hline 
\multicolumn{2}{|c|}{selection index}    &  B         &   C         &  D     \cr
\hline 
\end{tabular} \\
\end{center}
\caption{\capstyl Numbers of events with \lumi= 500\invfb\ expected both for signal and background processes at preselection level, standard selections (two set of cut on \brecoil) and multivariable analysis;  s/b and \fom\ are also indicated. }\label{res} 
\end{table} 

\begin{figure}[htbp]
\begin{center}
\begin{picture}(140,90)
\put(20,0){\epsfxsize110mm\epsfbox{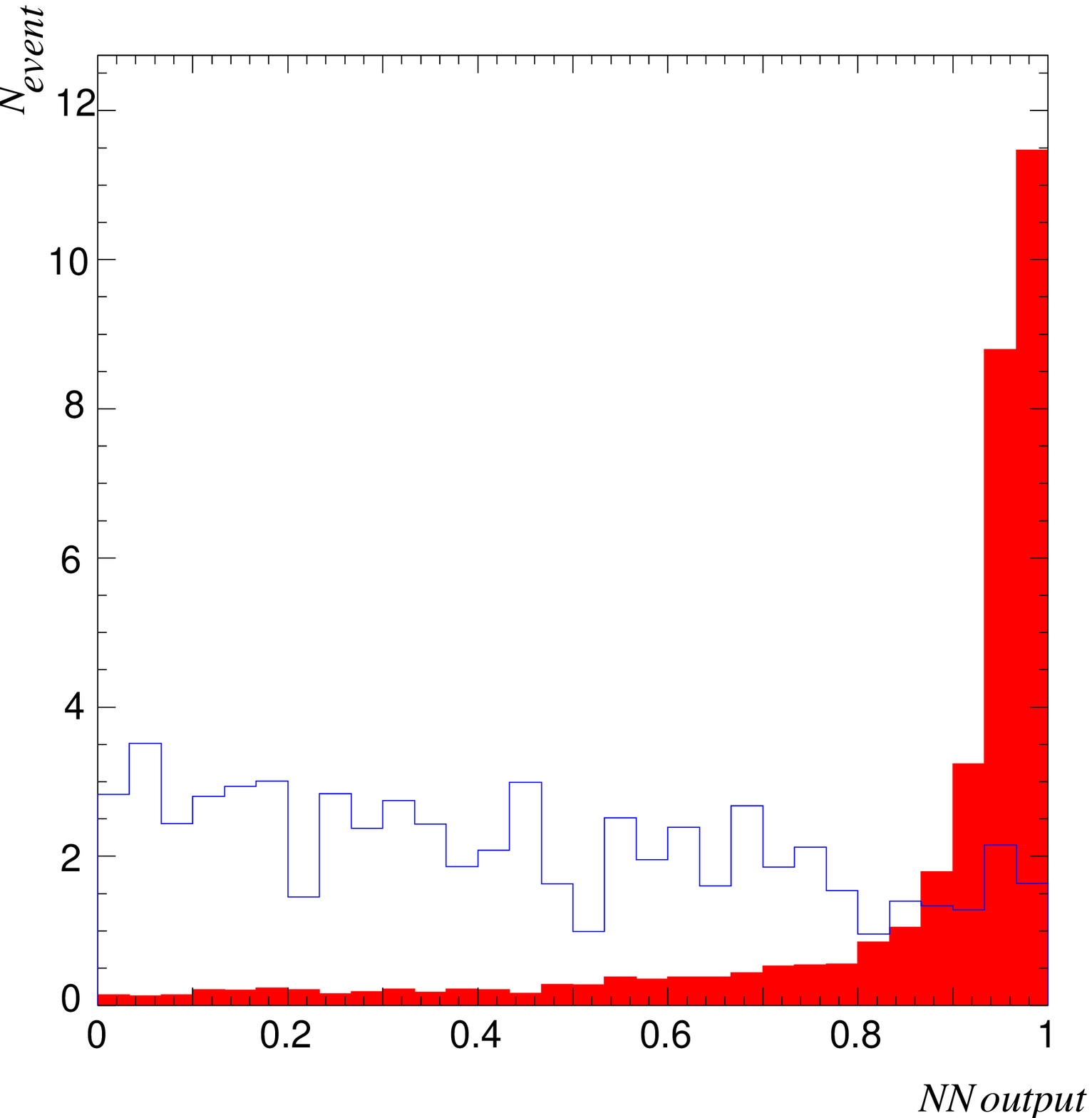}}
\end{picture}
\caption{\capstyl Neural Network output distribution ($ NN_{output}$) for $\rm H$$\rm H$$\rm Z$\ signal (full histogram) 
and background (empty histogram) with 500~{\mbox{$\hbox{fb}^{-1}$}}\ and 
{\mbox{$m_{\mathrm h}$}}=120~{\hbox{GeV}/\mbox{$\mathrm{c^2}$}}. 
\label{nn}}
\end{center}
\end{figure}

Clearly the combination of the the di-jets masses in the DIST variable
is a very crude one. Thus, in order to take into account the
correlation between the di-jets masses and the b content of the event,
the variables $m_{12}$, $m_{34}$, $m_{56}$ and \brecoil\ are combined
with a multivariable method. Here a neural network (NNet) approach has
been adopted. The NNet output varies from \mbox{-1}
 (background-like) to +1
(signal-like). Restricted to the positive values, Fig.~\ref{nn}
displays the NNet output from signal and background processes while
the contributions from the background sources are indicated in
Table~\ref{res}. The signal efficiency is about 36\% and s/$\rm
\sqrt{b}$ is $\sim$4.2 when an integrated luminosity of 500~\invfb\ is
assumed. The clear discrimination between the signal 
and the background processes could be exploited to reinforce the figure of merit~: as an example, with the same integrated luminosity, a cut on the NNet output at 0.5 leads to 31 signal events while 27 background events are
expected corresponding to a figure of merit of $\sim$~6. Better values 
could be obtained with thightened cuts and this behavior is 
fully used in the cross-section measurement where a fit of the 
NNet output distribution is performed as indicated in the next Section.

\section{Results}
\label{results} 
The cross-section of the \epem\goes\h\h\Z\ have to be measured, in order to extract \trilin. Such a measurement is performed by a likelihood fit of the characteristic variable (DIST, \brecoil, $\rm NN_{output}$) already used in the analysis. The different selections (indexed as B, C and D in Table~\ref{res}) have been tested. The relative error obtained on the cross-section measurement ($\Delta\sigma/\sigma$) according to the different sets of 'selection/discriminant variable' are reported in Table~\ref{sigma} for integrated luminosity from 500 to 2000~\invfb. A relative error of 10.3\% on the cross-section is obtained when NNet selection is used and an integrated luminosity of 2~\invab.

With the same analysis used for \mh=120~\Gcs, the relative error on the \epem\goes\h\h\Z\ cross-section has been evaluated for higher Higgs boson masses up to 140~\Gcs. The relative errors are reported in Table~\ref{masses} as well as the cross-section values and expected number of events with an integrated luminosty of 500~\invfb. Only the 'B/\brecoil' set of selection/variable has been considered. Due to the decrease of the cross-section the error increases rapidly up to 17\% even with an integrated luminosity of 2~\invab. Since this result is obtained with a standard selection, a dedicated analysis based on retrained NNet which take into account the Higgs boson mass would improve the performances.   

\begin{table}[h]
\begin{center} 
\begin{tabular}{|l|c|ccc|}
\hline
         selection & {variable}   & \multicolumn{3}{c|}{ $\Delta\sigma/\sigma$ } \cr
    &        & \lumi= 500\invfb & 1000\invfb  & 2000\invfb \cr 
\hline
  B &   DIST & 32.8 \%   &  25.6 \%         &   17.7  \%     \cr
  C &  DIST             &   29.8  \%        & 21.5 \%  &   15.1  \%    \cr
  B & \brecoil &  24.1 \%          &  17.3 \% &   11.6  \%              \cr
  D &   NN output        &     20.4  \%        &   12.9 \%            &    10.3  \%    \cr
\hline 
\end{tabular} \\
\end{center}
\caption{\capstyl Relative error ($\Delta\sigma/\sigma$) on $\sigma_{hhZ}$ for different selection/variable set and integrated luminosities} \label{sigma} 
\end{table}

\begin{table}[h]
\begin{center} 
\begin{tabular}{|l|c|c|c|ccc|}
\hline
\mh (\Gcs)  & $\rm \sigma_{hhZ}$(fb) & $\rm N_{hhZ}^{500}$ &  $\rm \epsilon_{hhZ}$ & \multicolumn{3}{c|}{ $\Delta\sigma/\sigma$ } \cr
 & & & & \lumi= 500\invfb & 1000\invfb  & 2000\invfb \cr 
\hline 
 120 & 0.186 &  93. &43\% & 24.1\%  &   17.3\% &  11.6\% \cr 
 130 & 0.149 &  74. &43\% & 26.6\%  &   19\%  &  17.7\% \cr
 140 & 0.115 &  57. &39\% & 32\%    &   23\%       &  17\%   \cr 
\hline
\end{tabular} 
\end{center}
\caption{\capstyl Relative error ($\Delta\sigma/\sigma$) on $\sigma_{hhZ}$ with the 'B/\brecoil' set for different Higgs boson masses and integrated luminosities; signal efficiencies ( $\rm \epsilon_{hhZ}$ ) and cross-sections ( $\rm \sigma_{hhZ}$ ) are reported and ($\rm N_{hhZ}^{500}$) the expected number of \h\h\Z\ events with \lumi=500\invfb} \label{masses}
\end{table}

\begin{figure}[htbp]
\begin{center}
\begin{picture}(140,70)
\put(10,0){\epsfxsize120mm\epsfbox{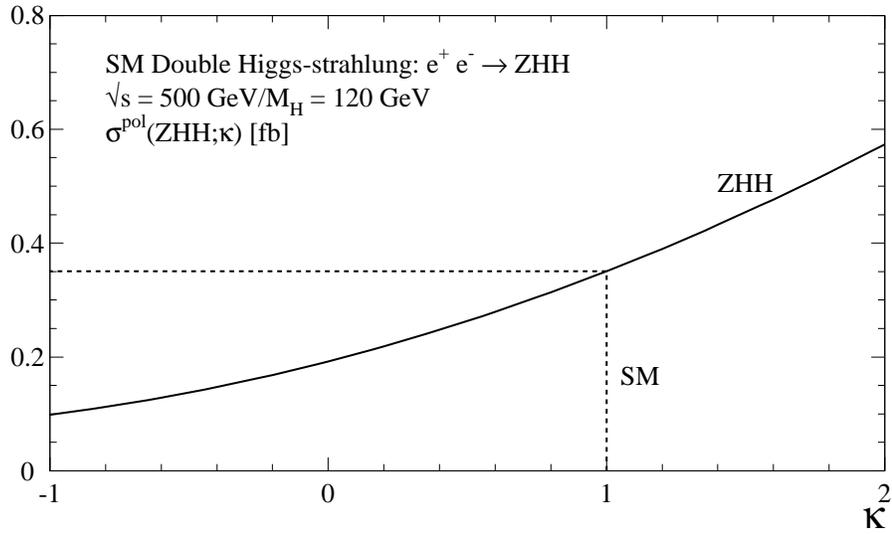}}
\end{picture}
\caption{\capstyl \epem\goes\h\h\Z\ cross-section (in fb) as a function of ${\kappa}$ when the incoming beams are fully polarized; \mh=120~\Gcs\ and \sqrtofs=500~GeV. 
\label{smvar}}
\end{center}
\end{figure}

To derive the relative error on \trilin\ ($\Delta\lambda/\lambda$) the
relation between \trilin\ and the cross-section has to be taken into
account according the behavior displayed on
Fig.~\ref{smvar}~\cite{maggie} where ${\kappa}$ is \trilin\ expressed
in unit of $\lambda_0$, the trilinear Higgs coupling in the \SM. Thus,
on this Fig.~\ref{smvar}, ${\kappa}$=1 stands for the \SM.
Furthermore, the incoming beams on that figure are assumed to be fully
polarized so that the displayed cross-section must be reduced by a factor
two~\cite{djouadi} to be coherent with the present unpolarized study.

As indicated by the Fig.~\ref{fda}, only one diagram for the
\epem\goes\h\h\Z\ cross-section is sensitive to \trilin. The others
constitute an irreducible background for the \trilin\ measurement
which shows up on Fig.~\ref{smvar} as the non-vanishing cross-section
at $\kappa=0$. Such an irreducible background makes the measurement
intrinsically difficult.

\begin{table}[h]
\begin{center} 
\begin{tabular}{|l|c|ccc|}
\hline
        {selection} &  {variable}     & \multicolumn{3}{c|}{ $\Delta\lambda/\lambda$ } \cr
  &        & \lumi= 500\invfb & 1000\invfb  & 2000\invfb \cr 
\hline
  B & \brecoil &  42.2\%& 30.3\% &  20.3  \%  \cr
  D &   NN output &  35.7\%&  22.6\%&    18.0\%    \cr 
\hline 
\end{tabular} \\
\end{center}
\caption{\capstyl Relative error ($\Delta\lambda/\lambda$) on $\lambda_{hhh}$ for different selections and integrated luminosities.} \label{tri}
\end{table}

From Fig.~\ref{smvar} the relative variations of the cross-section and
\trilin\ around the \SM\ are related by
$\Delta\lambda/\lambda$$\sim$1.75$\times \Delta\sigma/\sigma$.  The
relative error on \trilin\ is accordingly reported in Table~\ref{tri}
for two sets of selection/variable. With an integrated luminosity of
2~\invab\ and NNet output, the relative precision
$\Delta\lambda/\lambda$ is 18\%, and relaxes to 22\% with half of that
luminosity.

\section{Conclusions}
\label{conclusion} 
To establish the Higgs mechanism in an unambiguous way, the
self-energy potential of the Higgs field must be reconstructed. This
implies the determination of the trilinear self-coupling.  The
experimental feasibility of the $\lambda_{\rm hhh}$ measurement has
been explored trough a detailed analysis of the reconstruction of the
double Higgs-strahlung events. It takes advantage of the
characteristic signature with four b jets and a Z boson, reconstructed
either in its leptonic or hadronic decay modes. The large four and six
fermion background and the tiny signal cross section make of this
analysis a genuine experimental challenge. An excellent tagging as
well as reconstruction capabilities of the {\sc Tesla} detector are
then essential. Thanks to a multivariable method, based here on a
neural network (NNet), a $\rm s/\sqrt{b}$=6 could be achieved with 68
signal events selected in 1000~\invfb\ with a center-of-mass energy of
500~GeV and {\mbox{$m_{\mathrm
      h}$}}=120~{\hbox{GeV}/\mbox{$\mathrm{c^2}$}}; 18\% of the
selected signal events are from $\rm h$$\rm h${\mbox{$\ell^+\ell^-$}}.
With help of high integrated luminosity (2~\invab), the determination
of the {\mbox{$\mathrm{e^+ e^-}$}}{\mbox{$\rightarrow$}}$\rm h$$\rm
h$$\rm Z$\ cross-section with a 10\% relative error could be
achieved leading to a relative error on $\lambda_{\rm hhh}$ of 18\%.

\vspace{2cm} 
\subsection*{Acknowledgements} 
We gratefully acknowledge discussions with M. Battaglia, A. Djouadi, M. Muhlleitner and P. Zerwas.

\end{document}